\documentclass[%
reprint,
superscriptaddress,
nofootinbib,
amsmath,amssymb,
aps,
prl,
floatfix,
]{revtex4-1}

\usepackage{graphicx}% Include figure files
\usepackage{dcolumn}% Align table columns on decimal point
\usepackage{bm}% bold math
\usepackage{textcomp}
\usepackage[pdftex,colorlinks=true,bookmarks=false,citecolor=blue,urlcolor=blue]{hyperref}
\usepackage{cancel}
\newcolumntype{?}{!{\vrule width 1.1pt}}
\usepackage{layouts} % to determine textwidth in inches
\usepackage[normalem]{ulem}
\usepackage{xhfill}
\allowdisplaybreaks
\newcommand{\mrm}[1]{\mathrm{#1}}

\newcommand{\mum}{\ensuremath{\mathrm{\mu m}}}
\newcommand{\rvec}{\mathbf{r}}
\DeclareSymbolFont{lettersA}{U}{txmia}{m}{it}
\DeclareMathSymbol{\real}{\mathord}{lettersA}{"92} % real numbers
\DeclareMathSymbol{\cplx}{\mathord}{lettersA}{"83} % complex numbers

\newcommand{\subL}{\ensuremath{_{0,\mrm{L}}}}
\newcommand{\subS}{\ensuremath{_{0,\mrm{S}}}}

\newcommand{\ESedit}[1]{#1}
\newcommand{\ESeditAlt}[1]{#1}

\hypersetup{
   pdftitle=Laser wakefield driven generation of isolated CEP-tunable intense sub-cycle pulses
}

\begin{document}

\title{Laser wakefield driven generation of isolated CEP-tunable intense sub-cycle pulses}

\author{E.~Siminos}
\email{evangelos.siminos@physics.gu.se}
\affiliation{Department of Physics, University of Gothenburg, SE-412 96 G{\"o}teborg, Sweden}

\author{I.~Thiele}
\email{illia-thiele@web.de}
\affiliation{Department of Physics, Chalmers University of Technology, SE-412 96 G{\"o}teborg, Sweden}

\author{C.~Olofsson}
\affiliation{Department of Physics, Chalmers University of Technology, SE-412 96 G{\"o}teborg, Sweden}

\date{\today}

\begin{abstract}
Sources of intense, ultra-short electromagnetic pulses enable applications such as attosecond pulse generation, 
control of electron motion in solids and the observation of reaction dynamics at the electronic level.
For such applications both high-intensity and carrier envelope phase~(CEP) tunability are beneficial, yet hard to obtain with current methods.
In this work we present a new scheme for generation of isolated CEP-tunable intense sub-cycle pulses with 
central frequencies that range from the midinfrared to the ultraviolet. It utilizes an intense laser pulse which drives a wake in a plasma, 
co-propagating with a long-wavelength seed pulse. The moving electron density spike of the wake amplifies the seed and forms a sub-cycle pulse. 
Controlling the CEP of the seed pulse, or the delay between driver and seed leads to CEP-tunability, 
while frequency tunability can be achieved by adjusting the laser and plasma parameters. 
Our 2D and 3D Particle-In-Cell simulations predict laser-to-sub-cycle-pulse conversion efficiencies up to 1\%, resulting in relativistically intense sub-cycle 
pulses.
\end{abstract}

\maketitle

{Electromagnetic} pulses containing less than a single oscillation of the electromagnetic field are unique tools
for the investigation and exploitation of non-adiabatic phenomena.
One of the most prominent examples is the generation of attosecond pulses~\cite{Krausz14}, 
the efficiency of which depends 
on the carrier envelope phase~(CEP) of the driver pulse~\cite{Krausz107}. Moreover, an intense driver 
is particularly advantageous since it can produce even shorter (e.g. zeptosecond~\cite{PhysRevLett.111.033002}) pulses.
{Isolated sub-cycle pulses with both CEP-tunability and high energy} are {very} attractive not only for attosecond pulse generation, 
but also many other {applications} in solid-state physics~\cite{Gunter09,Hohenleutner15} and nano-engineering~\cite{Rybka16}. 
Therefore, developing methods to obtain isolated CEP-tunable high-intensity sub-cycle pulses is an active {field} of recent research. 
At present, {solid-state lasers {delivering relativistic intensity pulses} are limited to durations above one and a half cycle~\cite{Rivas17}, 
while scaling the intensity of sub-cycle pulses produced by parametric amplification methods remains challenging~\cite{Manzoni15}, 
especially in the mid-IR~\cite{Liang17}.} 

{Plasma based methods}, which are scalable even to relativistic intensities, offer hope to resolve this issue. 
{In the context of attosecond pulse generation in the XUV regime, several techniques have been developed in order to produce isolated pulses, such as polarization~\cite{baeva06,tzallas07,yeung15} or intensity gating~\cite{tsakiris06,heissler12,ma15,kormin18} and exploitation of wavefront rotation~\cite{Vincenti2012,wheeler12}. More recently, plasma-based methods have been proposed to generate longer wavelength, e.g.~mid-IR, single and sub-cycle pulses.}  
We have shown that electron beams can be used to generate intense sub-cycle pulses, by amplifying a seed pulse reflected by a foil~\cite{Thiele18}. 
Another technique {to generate midinfrared, near-single-cycle pulses,} which exploits the laser frequency down-conversion known to appear in laser-driven wakefields, has been proposed in \cite{nie2018}. However, {all {these plasma-based} techniques are either not CEP-tunable or require} a controllable CEP-stable high-intensity laser, which is {technically challenging}. 

\begin{figure*}
	\centering
	\includegraphics[width=1.0\textwidth]{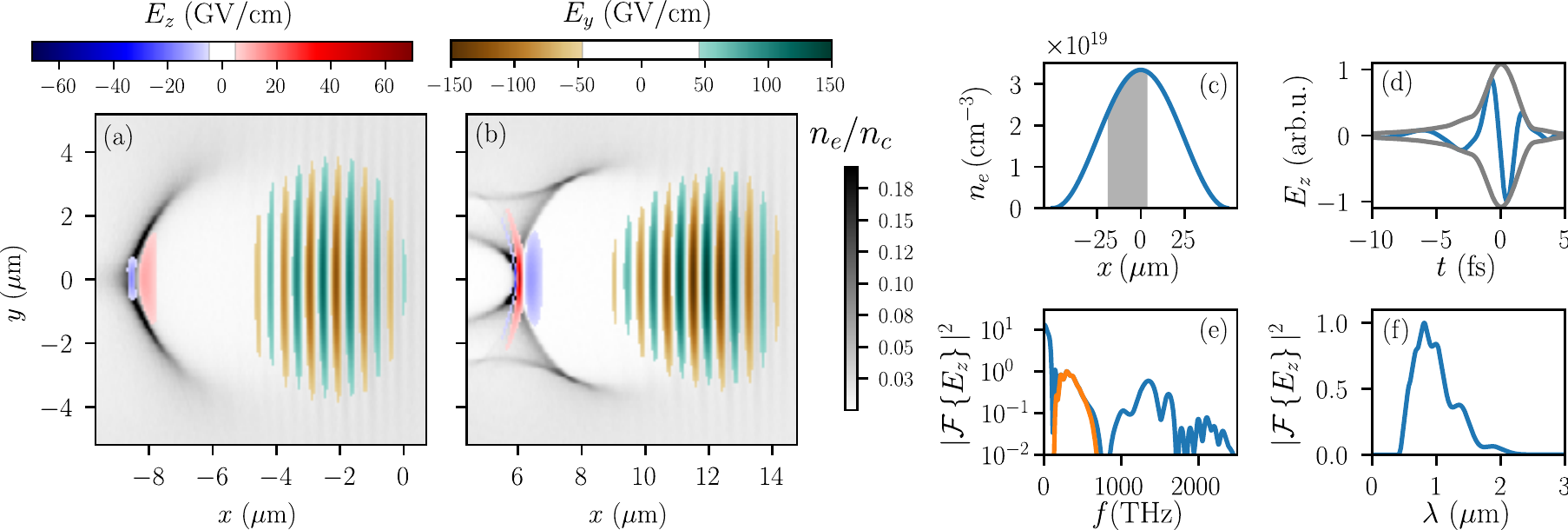}
	\caption{Results of a 3D particle-in-cell simulation demonstrating LWDA. Cross-sections of the plasma density and electromagnetic 
	field in the driver ($E_y$) and seed ($E_z$) polarization direction, (a) before and (b) after wavebreaking, (c) density profile used in the simulations, 
	with the grey-shaded area showing the region where amplification occurs, (d)~on-axis electric field of the sub-cycle pulse {at the exit of} the gas-jet, 
	(e)~power spectrum of the amplified pulse at the exit of the gas-jet, with the part of the spectrum retained after filtering shown in orange, and~(f) 
	the retained power spectra after filtering as a function of wavelength.}
	\label{fig:3D}
\end{figure*} 

In this article, we demonstrate that isolated CEP-tunable intense sub-cycle pulses can be created by a frequency up-conversion process, 
{which we refer to as} laser-wakefield driven amplification (LWDA). We propose to inject a CEP-stable long-wavelength seed pulse of relatively 
low intensity in co-propagation with a high-intensity, not necessarily CEP-stable, driver laser pulse into a gas jet. 
The driver pulse ionizes the gas already at its {rising edge} and creates a plasma. Due to the ponderomotive force, 
the high-intensity laser pulse~(driver) displaces the plasma electrons, creating a charge separation force which leads to plasma oscillations behind the laser pulse.
These wakefield oscillations~\cite{Gibbon,Es2009} {are} strongly anharmonic for sufficiently intense drivers and form electron density spikes after each oscillation period. 
In the highly non-linear regime, the first period is completely void of electrons (forming the so-called bubble) and the first electron density spike is most pronounced.
{In the scheme proposed here,} it is this electron density spike, moving close to the speed of light, which amplifies the seed pulse, leading to intense sub-cycle pulse generation.

We demonstrate sub-cycle pulse generation by means of a 3D PIC simulation~(see Fig.~\ref{fig:3D}) using the code SMILEI~\cite{SMILEI}. 
A $y$-polarized driving laser pulse and a $z$-polarized seed pulse are injected  along the $x$ direction. 
They are defined by their transverse electric field components $E_{y}$ and $E_{z}$, respectively, according to
\begin{equation}
E_{y/z}(\rvec_\perp,t)=E_{0,\mrm{L}/\mrm{S}}e^{-\frac{y^2+z^2}{r^2_{0,\mrm{L}/\mrm{S}}}}\sin\left(\phi_{\mrm{L}/\mrm{S}}(t)\right)f_{\mrm{L}/\mrm{S}}(t)\,\mrm{,}\label{eq:laser}
\end{equation}
with amplitudes $E_{0,\mrm{L}/\mrm{S}}$, beam waists $r_{0,\mrm{L}/\mrm{S}}$, angular frequencies $\omega_{0,\mrm{L}/\mrm{S}}$, 
phases  $\phi_{\mrm{L}/\mrm{S}}(t)=\omega_{0,\mrm{L}/\mrm{S}}t+\phi_{0,\mrm{L}/\mrm{S}}$, 
with variable seed CEP $\phi\subS$, {driver CEP} $\phi\subL=0$, and envelope of the form 
\begin{equation}
	f(t)=\exp\left(-2\ln(2)\ t^2/\,t_{0,\mathrm{L/S}}^2\right)\,\mrm{.}
\end{equation}
The simulation parameters are summarized in Table~\ref{tab:params}.
The gas-jet is modelled as a pre-ionized plasma with fixed ions and a cosine-squared electron density profile, see Fig.~\ref{fig:3D} 
with peak density $n_0=0.019\,n_c=3.3\cdot10^{19}$cm$^{-3}$, where $n_c=\epsilon_0\,m_e\omega\subL^2/q_e^2$ is the critical density and 
$q_e$ and $m_\mrm{e}$ are the electron charge and mass respectively.
The gas-jet has a short lengthscale with diameter (FWHM) $L_p=50\mum$, which is experimentally feasible~\cite{faure2018} (longer gas-jets can be used, as will be shown).
The driver laser has relativistic strength, characterized by normalized peak vector potential 
$a\subL={q_eE\subL}/{m_ec\omega\subL}=2.5$ (intensity $I\subL=1.3\times10^{19}\mathrm{W/cm^2}$). Its duration that satisfies $ct\subL<\lambda_{\mathrm{pe}}$, 
where $\lambda_{\mathrm{pe}} =2\pi c/\omega_{\mathrm{pe}}$ is the electron plasma wavelength and $\omega_{\mathrm{pe}}=\sqrt{n_0\,q_e^2/m_e\,\epsilon_0}$ 
is the plasma frequency. This ensures that a wake with a pronounced density spike in the first period is excited~\cite{Es2009}. 
The seed pulse wavelength has been chosen to correspond to $\lambda\subS=0.7\lambda_{\mathrm{pe}}=4\mum$ in order to
ensure propagation of the seed,
$\omega\subS>\omega_{\mathrm{pe}}$, while at the same time guaranteeing that
the seed wavelength is longer than the electron skin depth, 
$\lambda\subS>c/\omega_{\mathrm{pe}}=\lambda_{\mathrm{pe}}/2\pi$,
which can be taken as an \ESeditAlt{upper bound for} the density spike characteristic length scale~\cite{thomas2016}.
The seed is sub-relativistic ($a\subS=0.1$) while its energy is $\mathcal{E}\subS=0.3\mathrm{mJ}$, 
which is well within reach of optical parametric amplification~\cite{woodbury2018}.

% We consider a driver pulse with $\lambda\subL=800\,$nm central wavelength 
% with $60\,\mathrm{mJ}$, 800-nm driver pulse with duration $t_{0,L}=11\,\mathrm{fs}$ and waist $r_{0,L}=4.8$-$\mu$m. 
% This corresponds to peak intensity $I\subL=1.3\times10^{19}\mathrm{W/cm^2}$ 
% (peak electric field $E\subL=100$~GV/cm) or normalized peak vector potential $a\subL={q_eE\subL}/{m_ec\omega\subL}=2.5$, \emph{i.e.}, to a relativistic pulse.
% The seed pulse has central wavelength $\lambda_S=4\mum$, peak intensity $I\subS=8.6\times10^{14}\mathrm{W/cm^2}$ 
% (corresponding to field $E\subS=0.8$~GV/cm or $a\subS=0.1$), duration $t\subS=78$\,fs and $r\subS=27\mu$m. 
% This corresponds to $0.8\mathrm{mJ}$ of energy, which is well within reach of optical parametric amplification~\cite{woodbury2018}.  
% The simulations were run using a longitudinal resolution of 35 cells per (driver) wavelength, 
% transverse resolution of 15 cells per wavelength, 50 timesteps per optical cycle and 5 particles per cell. 
\begin{table*}[htbp!]
\begin{tabular}{|r||r|r|r||r|r|r|r||r|r|r|r|r||r|r|r|r|}
	\hline 	Fig.			&	$n_0$\,(cm$^{-3}$)	& 	$L_p$\,(\mum)	&	$\lambda_{pe}$\,(\mum)	& 	$\mathcal{E}\subL$ & a\subL	& $t\subL$\,(fs)	&$r\subL$\,(\mum)	&	$\lambda\subS$\,(\mum)	& $\mathcal{E}\subS$	&  $a\subS$ 	&	$t\subS$ 	& $r\subS\,(\mu$m) 	&	$N_x$	& $N_y$ 	& $N_t$ 	& $N_p$ \\ %& $a_{0,\mathrm{sub}}$ 	& $\lambda_{0,\mathrm{sub}}$ 	& $t_{0,\mathrm{sub}}$	
	\hline 	\ref{fig:3D}		&	$3.3\cdot10^{19}$	&	50		&	5.8			&	$60\,\mathrm{mJ}$  & 2.5	& $11$			& $4.8$			&	4			&	0.3\,mJ		&   0.1		&	78\,fs 		& 27			&	35	& 15		& 50		& 5	\\ %& 1.7			&	0.8\mum 		&	2\,fs
	\hline	\ref{fig:model}	&	$4.5\cdot10^{19}$	&	-		&	5			&	$39\,\mathrm{mJ}$  & 2.5	& $10$			& -			&	4 			& 	-		&   0.005	&	64\,fs		& -			&	105	& -		& 120		& - 	\\	
	\hline	\ref{fig:scheme}(a)	&	$4.5\cdot10^{19}$	&	23		&	5			&	$39\,\mathrm{mJ}$  & 2.5	& $10$			& $4.2$			&	4 			& 	-		&   0.005	&	CW		& 5			&	50	& 25		& 57		& 100 	\\
	\hline	\ref{fig:spec}(b,d) 	&	$1.75\cdot10^{18}$	&	152		&	25			&	$5\,\mathrm{J}$    & 2.5	& $50$			& $21$			&	20 			& 	-		&   0.005	&	CW		& 25			&	50	& 25		& 57		& 100 	\\
	\hline
\end{tabular}
 \caption{\label{tab:params}Parameters used in the simulations for the plasma (see text), driver and seed laser parameters in Eq.~\ref{eq:laser} (and also the 
 corresponding pulse energies $\mathcal{E}_{0,\mathrm{L/S}}$) and discretization parameters: longitudinal $N_x$ and transverse $N_y=N_z$ resolution in number of cells per (driver) wavelength, temporal  resolution $N_t$ in number of steps per optical cycle and number of particles per cell $N_p$.}
\end{table*}

% \color{green}{\rule[.5ex]{1em}{.4pt}\,\,\rule[.5ex]{1em}{.4pt}} 

Fig.~\ref{fig:3D}(a) shows how the first electron density spike of the wake behind the laser pulse amplifies the seed pulse and forms a sub-cycle pulse. The second and subsequent electron density spikes also create sub-cycle pulses. However, these have an intensity at least one order of magnitude smaller than that of the leading pulse, which is amplified by the dominant electron density spike, \ESeditAlt{resulting in an isolated sub-cycle pulse}. 
\ESeditAlt{Amplification occurs predominantly over a distance of approximately $20\mum$ in the rising edge of the gas-jet [grey-shaded area in Fig.~\ref{fig:3D}(c)], where the high  density gradient significantly enhances the leading density spike, while suppressing wavebreaking~\cite{Mu2013}.}
Amplification is interrupted shortly after the seed enters the declining part of the jet, due to wavebreaking associated with the plasma wavelength increase in the downramp~\cite{geddes2008}, see Fig.~\ref{fig:3D}(b).
The peak electric field of the seed is amplified by a factor of 84 from $0.8\,\mathrm{GV/cm}$ to $67\,\mathrm{GV/cm}$, while its peak wavelength is downshifted to $0.8\,\mum$ [see Fig.~\ref{fig:3D}(f)]. With respect to this wavelength the peak field of the sub-cycle pulse is relativistic, $a_{0,\mathrm{sub}}\simeq1.7$. The sub-cycle pulse is subsequently guided within the elongated plasma bubble until it exits the plasma, [\ref{fig:3D}(d)] with an ultra-broad spectrum, 
intensity FWHM duration of $\sim2\,$fs or 0.75-cycles and polarization orthogonal to the driver. The latter property allows the sub-cycle and driver pulses to be separated after the interaction. Most of the spectral power of the sub-cycle pulse lies within $\sim 2$ octaves from 0.45 to 2~\mum~[\ref{fig:3D}(f)]. \ESeditAlt{The energy of the amplified seed has increased by a factor of almost 3, reaching 0.86\,mJ, amounting to a pump to sub-cycle conversion efficiency of $\sim1\%$.}
\ESeditAlt{Note that spectral components outside the regime of interest for this sub-cycle pulse have been filtered out,}
% Note that high-frequency comonents associated to the interaction of the injected electron bunch with the seed as well as low-frequency 
%  components which are an artifact of measuring the fields in the near field in PIC simulations~\cite{dechard2019} have been filtered out, 
see Fig.~\ref{fig:3D}(e) and the Supplemental Material~\cite{suppl}. 

\begin{figure}
	\centering
	\includegraphics[width=\columnwidth]{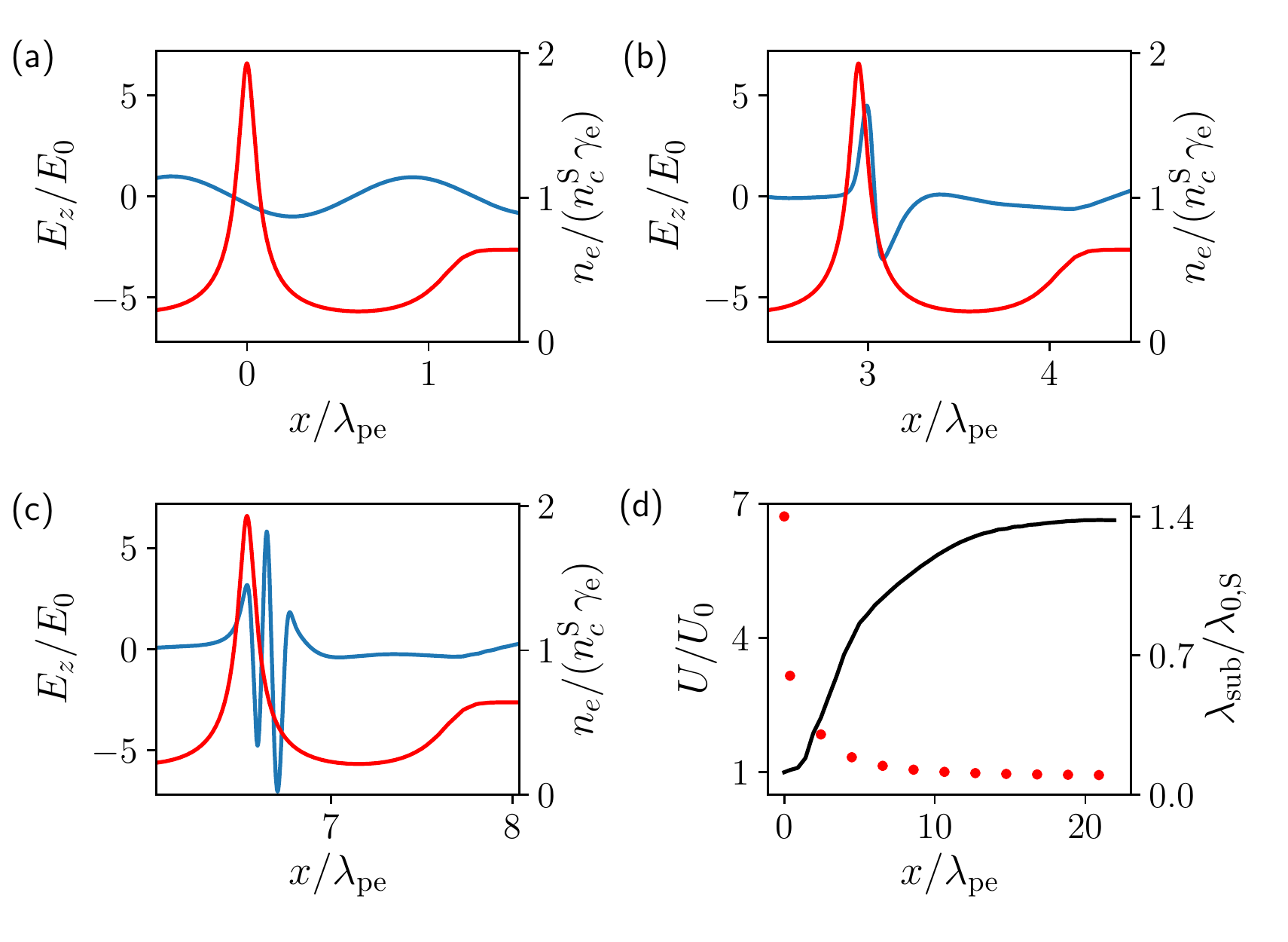}
	\caption{Amplification in the fluid model. (a-c)  seed electric field $E_z$, normalized to the initial seed amplitude $E_0$ (blue lines) and $n_e/\gamma_\mrm{e}=n_0/(1+\phi)$ from Eq.~(\ref{eq:phi}), normalized to the critical density $n_c^{\mathrm{S}}$ of the long-wavelength seed (red lines)  at different propagation distance.
	(d) Electromagnetic energy density $U$ in the polarization plane of the seed, normalized to the initial seed energy $U_0$ (black, solid line) and peak wavelength $\lambda_\mathrm{sub}$ of the sub-cycle pulse, normalized to the seed vacuum wavelength $\lambda\subS$ (red dots), as a function of propagation distance. For $x=0$, the wavelength corresponds to the seed peak wavelength in a homogeneous plasma. The parameters are given in Table~\ref{tab:params}.
	}
	\label{fig:model}
\end{figure}

\ESedit{In order to describe the amplification process in simpler terms, we develop a 1D model based on relativistic cold fluid theory. The propagation of the seed in the plasma can be described by the wave equation
\begin{equation}\label{eq:wave}
    \partial^2_{xx}A_z-\partial^2_{tt}A_z= n_e\,A_z/\gamma_\mrm{e}= n_0\,A_z/(1+\phi)\,, 
\end{equation}
where $\phi$, $n_e$ and $\gamma_e$ are the wake scalar potential, electron fluid density and relativistic factor, respectively, $A_z$ is the seed vector potential and $n_0$ the background plasma density, in relativistic units~\cite{suppl}. In the last step, use has been made of the fact that in the quasistatic approximation for the driver laser propagation $n_e/\gamma_\mrm{e}=n_0/(1+\phi)$~\cite{sprangle1990}. The scalar potential for the nonlinear wake is  determined by numerically solving~\cite{bulanov1989,sprangle1990,Es2009}
\begin{equation}\label{eq:phi}
    \frac{d^2\phi}{d\xi^2} = n_0\left(\frac{1+A_y^2}{2(1+\phi)^2}-\frac{1}{2}\right)\,,
\end{equation}
where $A_y$ is the vector potential of the driver laser, $\xi=x-v_g\,t$ and $v_g\simeq c\,(1-1.5\,n_0/n_c)$ is the driver group velocity~\cite{lu2007}. The first period of the steady state, 
wakefield solution propagating at $v_g$ is shown in Fig.~\ref{fig:model}(a).
%, in units of the critical density $n_c^{\mathrm{S}}$ of the long-wavelength seed. 
The solution of Eq.~(\ref{eq:phi}) for $\phi$
is used in order to solve Eq.~(\ref{eq:wave}) numerically.
%While Eq.~\ref{eq:A_z} cannot be solved analytically with the nonlinear source term obtained from Eq.~\ref{eq:phi}, we can obtain useful information by numerically solving it for $A_z$, with the solution shown in Fig.~\ref{1d}(a).
As shown in Fig~\ref{fig:model}(d), the model predicts rapid seed wavelength decrease, as well as significant electromagnetic field energy gain. Fig.~\ref{fig:model}(b) shows that the wavelength downshift and energy gain is associated to localized amplification at the front of the density spike of the wake. This is consistent with the prediction of Ref.~\cite{Thiele18}, that a pulse can gain energy from the declining part (with respect to $x$) of a sub-wavelength, moving density perturbation. This effect is distinct from the frequency upshift or ``photon acceleration''~\cite{Wilks89,Esarey90,Silva98,Murphy06,Bu15}, associated with electromagnetic field propagation in a plasma with (spatially) decreasing density. In the latter case the pulse wavelength is shorter than the density variation length-scale \ESeditAlt{and pulse energy increase is only possible through an associated pulse length increase~\cite{Esarey90}}. 
\ESeditAlt{Nevertheless, also in our case plasma propagation effects are important and a purely beam-driven description~\cite{Thiele18} does not apply.}}
Moreover, the amplification process is not due to a parametric process such as stimulated Raman scattering~\cite{kruer}, since the driver and seed polarization are orthogonal. 
Finally, note that LWDA is distinct from the formation of optical bullets 
in laser-wakefield acceleration, which occurs when a \ESeditAlt{short wavelength} probe is trapped inside the plasma density depression (bubble) created by the driver laser~\cite{sheng2000,dong10}.

\ESedit{After propagation for $\sim5\,\lambda_{\mathrm{pe}}$, the rate of  energy increase starts to decrease, since the local seed wavelength is not anymore longer than the density modulation length-scale. Moreover, the increased group velocity, due to the decrease in wavelength, of the amplified seed implies that the latter starts to dephase with respect to the density spike, see Fig.~\ref{fig:model}(c), leading to a gradual pulse lengthening. 
\ESeditAlt{In the 3D PIC simulations of Fig.~\ref{fig:3D} wavebreaking occurs before dephasing becomes an issue and the sub-cycle pulse is guided within the bubble until the gas-jet exit, allowing the use of gas-jets of reasonable dimensions. We note that \ESeditAlt{while the static wakefield model captures very well the salient features of sub-cycle pulse formation, amplification} is higher in PIC simulations, due to the additional enhancement of the first density spike induced by the driver laser evolution in that case.} 
%Nevertheless, this simple model captures very well the salient features of formation of sub-cycle pulses in PIC simulations, in particular the energy gain and frequency shift, as well as the CEP-tunability which will be discussed below. 

%Two-dimensional PIC simulation with the same interaction conditions show that the interaction is terminated by 
%In PIC simulations, in particular in multi-dimensional ones, such as the one presented in Fig.~\ref{fig:3D}, the process is much more efficient due to  the first density spike for the same interaction conditions 

%is much more pronounced, due to the evolution of the wake leading to greater  practice, longer gas jets can be used as shown in the example of the 3D PIC simulation

%We note that as a result of the difference in gradient of $n_e/\gamma$ in the middle of the wake and in the density spike, the pulse is chirped, as also observed in PIC simulations.
}

\begin{figure}[hbtp!]
	\centering
	\includegraphics[width=\columnwidth]{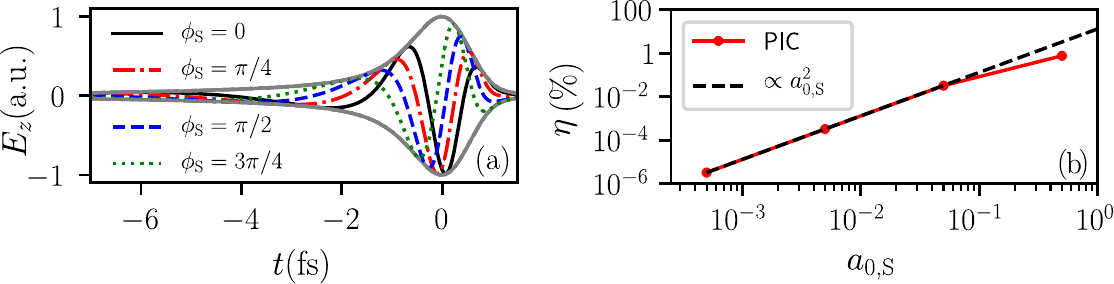} 
	\caption{Results of 2D PIC simulations.~(a)~on-axis electric fields of the amplified sub-cycle pulses after the interaction for different seed delays $t_d$ and their common envelope~(gray line), (b)~energy conversion efficiency {dependence on the peak} seed electric normalized potential. 
	}
	\label{fig:scheme}
\end{figure}

To discuss some additional features of the interaction we perform a parametric study with 2D simulations in a simplified setting in order to reduce computational costs, while still capturing the main effects (see~\cite{suppl} for a comparison of 2D and 3D results).
We consider a flat-top plasma density profile, the seed pulse is modeled within the continuous wave approximation (CW), $f_S(t)=1$ in Eq.~(\ref{eq:laser})
and the remaining simulation parameters are summarized in Table~\ref{tab:params}.
%  As a reference case we take a simulation with a flat-top density profile with $n_0=0.0256\,n_c$, driver laser with $a\subL=2.5$, $t\subL=10\,$fs, $r\subL=5\mum$ 
%  and seed pulse with $\lambda_S=4\mum$, $r\subS=4.2\mum$, $a\subS=0.005$, modeled 
%  In 2D we use 100 macro-particles per cell, longitudinal (transverse) resolution of 50 (25) cells per driver wavelength and temporal 
%  resolution of 57 steps per optical cycle. 
Fig.~\ref{fig:scheme}(a) shows the generated sub-cycle pulse at the exit of the plasma with 
 half-cycle duration and peak wavelength of $\lambda_{\mathrm{sub}}=0.4\,\mum$.
 
In order to tune the CEP of the sub-cycle pulse, it is sufficient to introduce a delay, $t_d$, of the many-cycle seed pulse with respect to the laser pulse or to change the CEP of the seed $\phi_{\mathrm{S}}$. 
In the CW approximation studied here these two operations are equivalent, since $t_d\omega_{0,\mrm{S}}=\phi\subS$; the finite pulse duration case is studied in~\cite{suppl}. 
Figure~\ref{fig:scheme}(a) presents the on-axis electric field shapes after the interaction using four different seed pulse delays in steps of $\lambda\subS/8$ (or phase-shifts of $\pi/4$).  %CEPs $\phi_S$
It can be seen that the sub-cycle pulse envelopes are the same, %as are the power spectra of the pulses~[see Fig.~\ref{fig:scheme}(d)]. However, 
however the phases of the sub-cycle pulses are shifted by a value of $\pi/4$. 
The synchronization level necessary  to achieve a change of $\pi/2$ in CEP is $\lambda\subS/4$. For the range of densities studied here, this corresponds to $3-17\,$fs, which is within present experimental capabilities~\cite{golovin2018}. {Moreover, we show in~\cite{suppl} that the sub-cycle phase is not sensitive to variations in the driver laser {pulse} duration or energy by up to 20\%, making such control strategy viable.}

%According to Fig.~\ref{fig:scheme}(b), {which shows} the sub-cycle pulse energy $\mathcal{E}_\mrm{out}$ as a function of the gas-jet length $L_\mrm{p}$, the amplification takes place during the first 30~$\mu$m of propagation.
%For longer propagation, relativistic self-focusing of the laser pulse leads to bubble lengthening and wavebreaking~\cite{thomas2007,kalmykov2009,savert2015} and amplification ceases. 
%After wavebreaking the sub-cycle pulse is guided inside the bubble for distances of up to the driver depletion length $L_{\mathrm{pd}}\propto a_\mathrm{0,L}\lambda_\mathrm{pe}^3/\lambda\subL$. 
%For the present parameters, our simulations show that the driver laser is able to drive a bubble for about $200\,\mum$ of propagation. 

%Figures \ref{fig:scheme}(c,d), show the sub-cycle field strength scaling with the seed pulse duration and beam waist $r_{0,S}$. We see that amplification remains efficient over a wide range of values of both seed pulse duration and waist. This eases temporal and spatial overlap requirements, since one can work with a relatively long and wide seed. The main requirement on the seed is that its wavelength is longer, and its focal spot wider, than the characteristic length $\sim c/\omega_{\mathrm{pe}}$ of the density spike. 

According to the cold fluid model, Eq.~\ref{eq:wave}, the interaction is linear in the seed transverse vector potential, implying a 
%linear scaling of the sub-cycle electric field amplitude and
quadratic scaling of the energy conversion efficiency with the seed electric field amplitude. This is confirmed by our 2D PIC simulation results in Fig.~\ref{fig:scheme}(b),up to weakly relativistic seed pulse amplitudes. 
% To judge how relativistic the seed and sub-cycle pulses are, % the field strength in this figure is given in terms of the normalized vector potential $a\subS$ and $a_{\mathrm{sub}}$, respectively (calculated using the corresponding central wavelengths $\lambda_\mathrm{S}=4\mum$ and $\lambda_\mathrm{S}=0.4\mum$). 
The sub-optimal scaling for $a\subS>0.1$ is caused by the \ESeditAlt{feedback of the seed pulse on the wake. In this case, the undepleted driver approximation, implicit in Eq.~(\ref{eq:phi}), does not hold; the wake loses a substantial fraction of its energy to the seed pulse and $\phi$ in Eq.~(\ref{eq:wave}) cannot be considered to be stationary and independent of the sub-cycle field $A_z$ anymore}. Note that, as discussed in relation to Fig.~\ref{fig:3D}, relativistic field strengths can nevertheless be reached using sub-relativistic seed pulses.
Figure~\ref{fig:scheme}(b) shows that the laser-to-sub-cycle-pulse conversion efficiency $\eta$ reaches about 1\%. 

\begin{figure}
	\centering
	\includegraphics[width=\columnwidth]{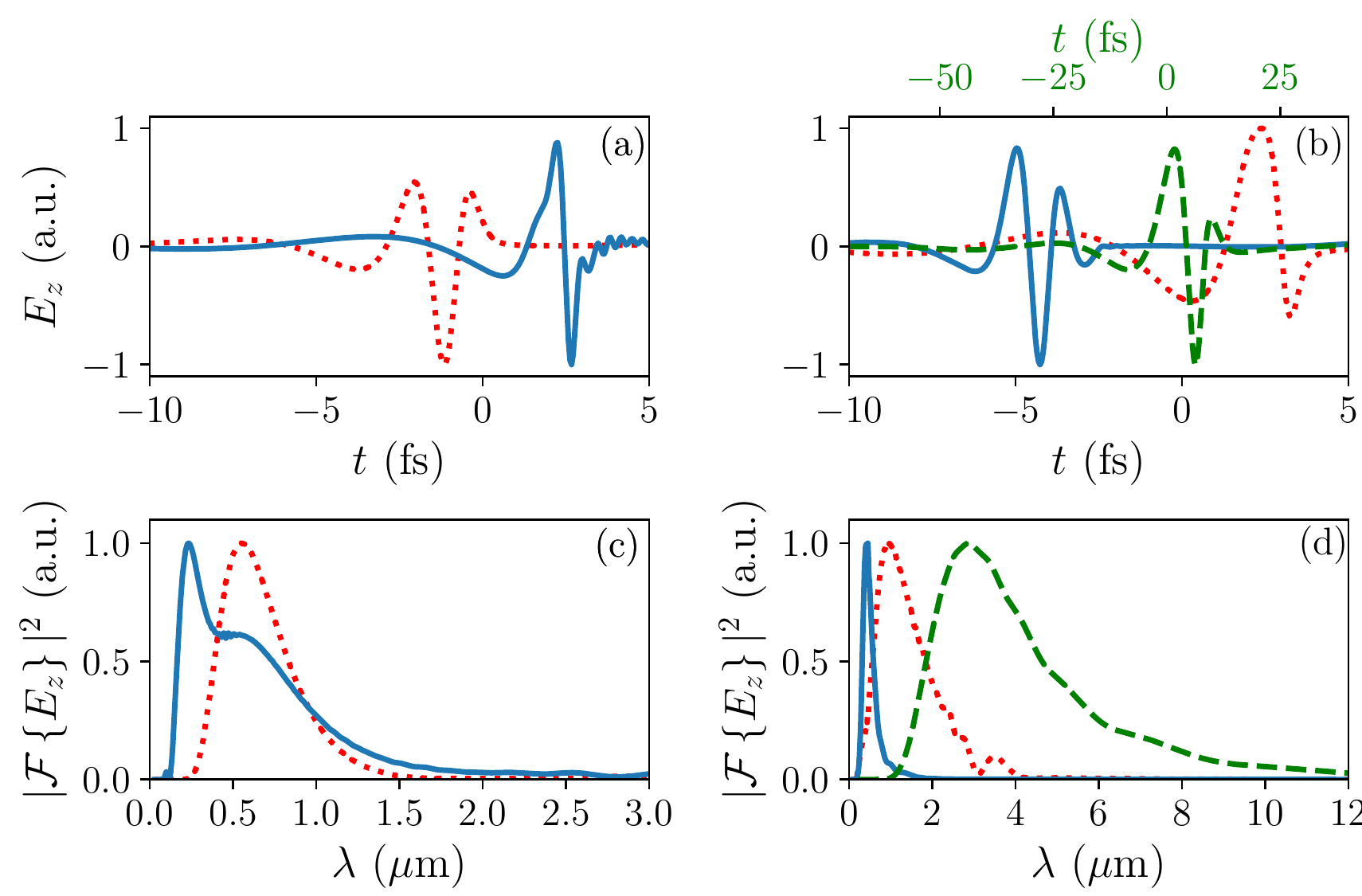}
	\caption{Demonstration of the spectral tunability by modifying (a,c)~laser peak normalized vector potential from $a_\mathrm{0,L}=2$ (red dotted lines) 
	to $a_\mathrm{0,L}=3.5$ (blue solid lines) and (b,d)~electron density from $n_\mrm{0}=0.019\,n_c=3.35\cdot10^{19}$~cm$^{-3}$~(red dotted lines) 
	to $n_\mrm{0}=0.045\,n_c=7.8\cdot10^{19}$~cm$^{-3}$~(blue solid lines), keeping the other parameters the same as in Fig.~\ref{fig:scheme}. 
	For the green dashed lines in (b,d) the density has been reduced to $n_\mrm{0}=0.001\,n_c=1.75\cdot10^{18}$~cm$^{-3}$, 
	while the rest of the parameters have been rescaled in proportion to $\lambda_{\mathrm{pe}}$ (see Table~\ref{tab:params}). 
	Note the different time-scale in this case.
	The figures present the (a,b)~on-axis electric fields of the sub-cycle pulses
	after the interaction as well as (c,d)~their power spectra.}
	\label{fig:spec}
\end{figure} 

% spectral tunability
\ESedit{The 1D model suggests that, apart from the interaction length,} the sub-cycle pulses' spectra could be tuned with the driver-laser field strength $a_\mrm{0,L}$ and electron density $n_\mrm{0}$. 
The larger $a_\mrm{0,L}$ and $n_\mrm{0}$ are, the narrower the amplifying electron density spikes, 
which naturally leads to shorter wavelength components in the sub-cycle pulse spectra, see Fig.~\ref{fig:spec}. 
% Both the central wavelength and the wavelength range, can be tuned as shown in Fig.~\ref{fig:spec}.  
%{By using} more intense driver laser pulses, we can {introduce shorter wavelength components in} the spectrum, see Fig.~\ref{fig:spec}(a,c).\ES{repetitive} 
Tuning through the use of $a_\mrm{0,L}$ is limited by early wave-breaking of the amplifying electron density spike above $a_\mrm{0,L}=3.5$ (for this particular plasma density). 
%Reducing the electron density (keeping all other parameters fixed) shifts the sub-cycle pulse peak wavelength towards the midinfrared, 
%while increasing the density shifts it towards the ultraviolet, see Fig.~\ref{fig:spec}(b,d). 
{Tuning through the variation of $n_\mrm{0}$ is, according to our simulations, effective as long as $0.7\,\lambda_{\mathrm{pe}}\lesssim\lambda\subS\lesssim\lambda_{\mathrm{pe}}$}.

In order to scale our results to a wider range of densities, one can fix 
the driver laser wavelength to $\lambda_L=0.8~\mu$m and vary $n_0/n_c$, while  
scaling all other parameters in proportion to the plasma wavelength.
% In particular both the driver and seed laser pulse duration and waist are scaled in proportion to $\lambda_{\mathrm{pe}}$, % while we take $\lambda\subS=0.8\lambda_{\mathrm{pe}}$ for all cases. 
An example for low density plasma, $n_0=0.001\,n_c=1.7\cdot10^{18}\mathrm{cm}^{-3}$, is shown as a green dashed line in   
 Fig.~\ref{fig:spec}(b,d), see Table~\ref{tab:params}. Due to the longer seed wavelength used in this case, 
 $\lambda\subS=20\mum$, the generated sub-cycle pulse lies in the midinfrared.

In summary, we propose a scheme for the generation of isolated, CEP-tunable relativistic sub-cycle pulses {by} 
laser wakefield driven amplification of a seed electromagnetic pulse. 
The scheme has been shown to work over a wider range of plasma densities, utilizing seed pulses with wavelengths 
ranging from $4\,\mum$ to $20\,\mum$ to produce relativistic sub-cycle pulses with peak wavelengths adjustable from the midinfrared to the ultraviolet. 
% While here we have focused to the simulation of relatively short plasmas, ranging from few tens two few hundreds of microns, resulting in field amplification factors between 10 and 90, 
% While there is room for further optimization of the method, for example by extending the propagation length by suppressing wavebreaking,
% generation of relativistic sub-cycle pulses is within reach already at the conversion efficiency of the order of 1\% reported here. 

\begin{acknowledgments}
  The authors thank J.~Ferri, M.~Grech, L.~Gremillet, G.~Golovin, D.~Gu\'{e}not, V. Horn\'{y}, M.~Kaluza and L.~Veisz for helpful suggestions, T.~Blackburn for a careful reading of the manuscript and T.~F\"ul\"op for support.
  This work was supported by the Knut and Alice Wallenberg
  Foundation and by the Swedish Research Council, Grant No. 2016-05012. Numerical simulations were performed using computing resources at Grand {\'E}quipement National pour le Calcul Intensif (GENCI, Grants No.~A0030506129 and No.~A0040507594) and Chalmers Centre for Computational Science and Engineering (C3SE) provided by the Swedish National Infrastructure for Computing (SNIC, Grants 2017/1-484, 2017/1-393, 2018/3-297, 2018/2-13).
\end{acknowledgments}

% \bibliography{mybibfile}

%merlin.mbs apsrev4-1.bst 2010-07-25 4.21a (PWD, AO, DPC) hacked
%Control: key (0)
%Control: author (8) initials jnrlst
%Control: editor formatted (1) identically to author
%Control: production of article title (-1) disabled
%Control: page (0) single
%Control: year (1) truncated
%Control: production of eprint (0) enabled
%

% \clearpage

% \input Supp_Matt

\end{document}